# Treatment of PDMS surfaces using pulsed DBD plasmas: comparing the use of different gases and its influence on adhesion

Fellype do Nascimento[1,a], Stanislav Moshkalev[1], Munemasa Machida[2], Sergio Parada[1]

[1] *Center for Semiconductor Components – State University of Campinas*

[2] *Instituto de Física "Gleb Wataghin" – State University of Campinas*

## Abstract

In this work we present some results of the treatment of polydimethylsiloxane (PDMS) surfaces using pulsed dielectric barrier discharge plasmas. The results of plasma treatment using different gases and mixtures, argon, argon plus water vapor, helium, helium plus water vapor, nitrogen and nitrogen plus water vapor, were compared testing the adhesion between two PDMS samples for each kind of plasma. We also studied the water contact angle in function of plasma process time of PDMS surfaces with each kind of plasma treatment. The plasma was characterized by optical emission spectroscopy (OES) to identify the emitting species and determine the plasma temperatures. The plasma temperature for each process was estimated comparing the spectrum obtained by OES with the spectrum generated by SpecAir simulation code. Measurements of power delivered to the plasmas were also performed. As the results, all the process using different gases show good adhesion efficacy between PDMS samples when long exposure time (larger than 150 seconds) is applied. However, when only a few discharges are applied to PDMS samples the helium plasma process presented best results. Atomic Force Microscopy (AFM) analysis of PDMS samples treated with helium plasma showed reduction in the surface roughness, which increase the surface contact area and improves the adhesion.

## Introduction

Dielectric Barrier Discharge (DBD) plasma is a non-equilibrium plasma that can be generated at atmospheric pressure, both in open or closed enviroment. In this kind of plasma the discharges are produced between two electrodes with at least one of then covered with a dielectric material (glass or ceramic in most cases). In an open atmosphere the plasma discharges can be produced with a gas flow between the electrodes. One attractive characteristic of the DBD plasma at atmospheric pressure is that it is a low temperature plasma [Bibinov2001,Masoud2005,Rajasekaran2012,Bashir2014,Machida2015] and, for this reason, can be used to modify or activate surfaces of a wide range of materials, from polymers to biological tissues, without damaging them. To generate the DBD plasma we used a device that was built using a 5C22 thyratron valve and ferrite transformer [Machida2015]. The device can be operated from 5 to 40kV and can produce from 6 to 300 plasma pulses per second.

---

[a] Corresponding author: fellype@gmail.com

Polymers like polydimethylsiloxane (PDMS) have numerous applications. Depending on the application, a treatment on the surface of the PDMS is required. The use of plasma sources for surface treatment is a very known technique [Lommatzsch2007,Chiang2010,Yang2013,Kostov2014]. The treatment of polymers materials can be done using radio-frequency produced oxygen plasmas [Bodas2007,Xiong2014] in a vacuum chamber, radio-frequency plasmas at atmospheric pressure [Lommatzsch2007], alternate current DBD plasmas [Shao2011] or using a pulsed DBD plasma [Yang2013]. In this work we present some results of the treatment of PDMS surfaces using pulsed dielectric barrier discharge (PDBD) plasmas. The adhesion between PDMS surfaces treated with PDBD plasmas generated using flows of six different gases, argon, argon plus water vapor, helium, helium plus water vapor, nitrogen and nitrogen plus water vapor, was annalized considering adhesion efficacy, emission spectroscopy and power delivered to the plasmas. Since there is a correlation between water contact angle (WCA) and adhesion properties, where lower contact angle results in better adhesion [Gadre2003,Lommatzsch2007,Chiang2010], we studied the WCA in function of surface treatment time for all the six different plasmas. The pressure supported in adhesion between two PDMS samples were also determined in all cases using only a few plasma discharges to treat the PDMS surfaces. Atomic Force Microscopy (AFM) analysis of PDMS samples treated with helium plasma showed reductions in the surface roughness even when only a few plasma pulses are applied to the PDMS surface. Similar changes in the surface roughness were also observed for other polymeric materials treated with plasmas [Noeske2004,Nastuta2008,Slepicka2013,Kostov2014]. We also observed that the surface roughness of a PDMS sample does not present significative change with aging time.

The characterization of the plasmas used in this work were made using optical emission spectroscopy (OES) and power measurements. OES was used in order to know what reactive species are present in the plasmas and to determine rotational and vibrational temperatures ($T_{rot}$ and $T_{vib}$, respectively). Since the gas temperature ($T_{gas}$) has approximately the same value as the rotational temperature of diatomic molecules in non-equilibrium plasmas [Motret2000,Moon2003,Bruggeman2014], a comparison between experimental and simulated $N_2$ I emission lines were performed to estimate $T_{gas}$ in the characterization of DBD plasmas.

**Experimental setup**

Figure 1 shows the scheme of the apparatus used to treat PDMS samples with DBD plasma. The part consisting of tubes and electrode is a transverse section of the DBD reactor.

The device operates according to the following: a continuous gas flow is injected inside the PVC tube and a high-voltage pulse is applied to the electrode inside the glass tube. A primary discharge is formed in the region between the glass tube and the PVC tube producing a plasma jet leaving the tube exit which is used for surface treatment of PDMS samples.

In order to produce plasma discharges of some gas with water vapor, the water vapor is carried with the gas flow via the bubbling method. The vessel with water indicated in the dashed part of Fig. 1 is present only when we want to produce plasma of a gas with water vapor. In all other cases, the tube leaving the flowmeter is plugged directly to PVC tube. The use of water vapor was intended to increase the number of OH molecules in the plasma [Sarani2010]. To make the surface treatment all the PDMS samples were placed at a distance $d \approx 3.0$mm from the end of PVC tube.

Measurements of spectral emissions were made using an Andor 303i spectrometer equipped with an iStar DH720 iCCD detector. A 150 lines/mm grating was used in order to get an overview of entire spectra and a 1200 lines/mm grating was used for detailed spectral measurements. The light emitted by the plasmas were collected with a lens and transported to the spectrometer through an optical fiber.

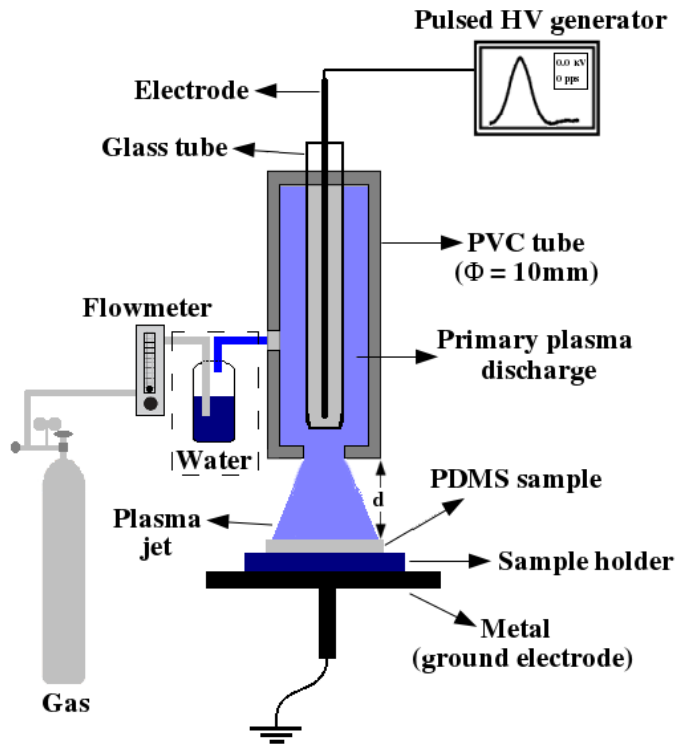

**Figure 1:** Scheme of the apparatus for treatment of PDMS samples with DBD plasma. The DBD reactor (PVC tube, glass tube and electrode) is shown in a transverse section. Φ refers to internal diameter of PVC tube. The elements are out of scale.

# Results and discussion

*Plasma characterization*

**– Optical emission spectroscopy**

The emission spectra of the plasmas made with argon, argon plus water vapor, helium, helium plus water vapor, nitrogen and nitrogen plus water vapor gas flows are shown in Figure 2. The spectrums in Fig. 2 are accumulations of 1500 plasma pulses acquired using a 150 lines/mm grating. The same setup used to treat PDMS samples (Fig. 1) were used to acquire the spectrums.

In Fig. 2 we can see that OH I molecular line emissions (near 308nm) are present in the emission spectra of all plasmas. Some $N_2$ I molecular line emissions are also present in the emission spectra of all plasmas. The most intense $N_2$ I line emissions appears at 337.1, 357.7 and 380.5 nm. $N_2$ II molecular line emissions occurs only for helium plasma (at 391.4 and 427.8 nm). An atomic line emission from oxygen (O I at 777.2 nm) was observed for argon and argon plus water plasmas only.

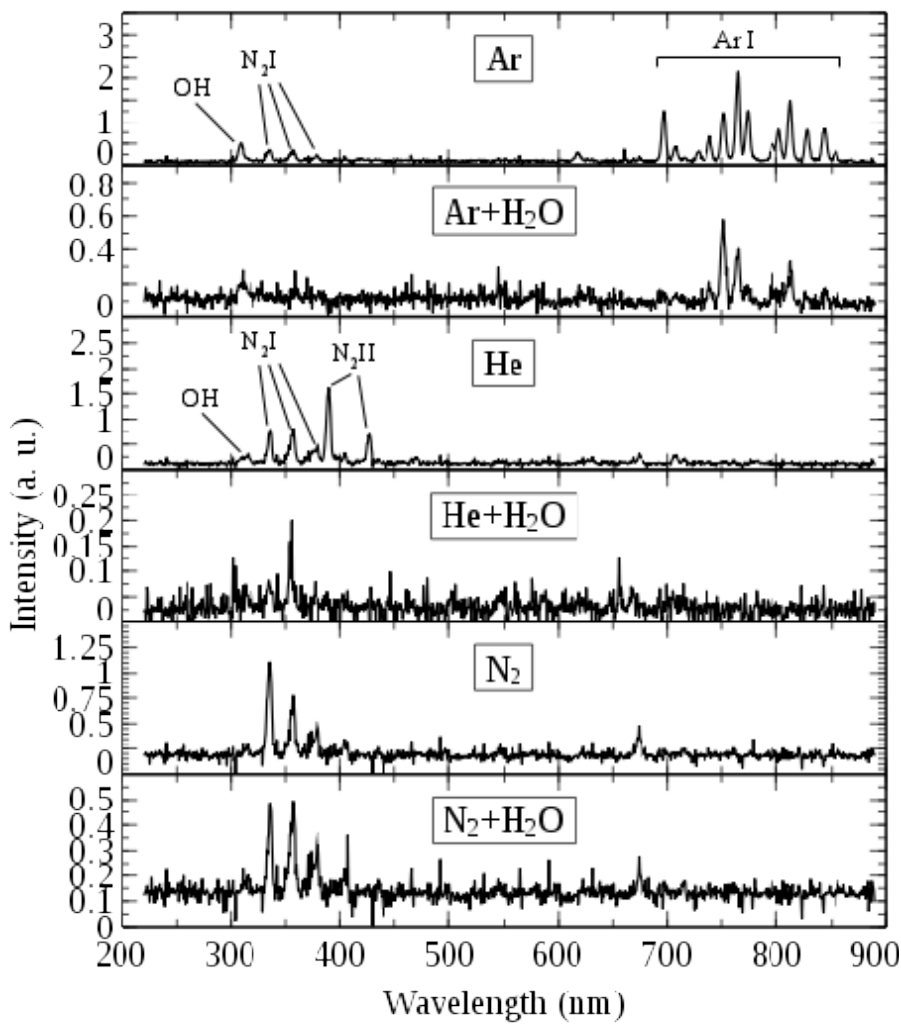

**Figure 2:** Emission spectrums obtained for plasmas of argon, argon plus water vapor, helium, helium plus water vapor, nitrogen and nitrogen plus water vapor gases.

The emission spectra of argon and argon plus water plasmas shows many atomic argon emission lines (Ar I at the end of spectrums). When using the helium and helium plus water gases, there is an atomic helium line emission (He I at 587.6 nm), with very weak intensity, only for helium plasma.

As we can observe in Fig. 2, the use of water vapor reduces the intensity of some spectral emissions and suppress others. This indicates a reduction in the power delivered to the plasma when using water vapor in a gas flow to generate the plasma. We can also observe that the use of water vapor did not increase the intensity of OH light emission like ocurred in [Sarani2010].

Figure 3 (a-d) shows $N_2$ I emission lines (second positive, $C^3\Pi_u - B^3\Pi_g$ transitions) in the wavelength range from 365 to 385 nm of plasmas generated using four different gases: argon, helium, nitrogen and nitrogen plus water vapor. The comparison between experimental and simulated curves were used to estimate the rotational temperature, that is approximately equals to the gas temperature ($T_{rot} \approx T_{gas}$) [Motret2000,Moon2003,Bruggeman2014], and the vibrational temperature ($T_{vib}$) for nitrogen molecules in each case. The simulations were carried out using SpecAir software [refSpecAir]. Table 1 Summarizes the results obtained for $T_{rot}$ and $T_{vib}$. We assume an uncertainty of 10% on each temperature value.

**Table 1:** Rotational and vibrational temperatures for $N_2$ for each kind of plasma

| Gas | $T_{rot} \approx T_{gas}$(K) | $T_{vib}$ (K) |
|---|---|---|
| Ar | 550 | 1800 |
| He | 400 | 2800 |
| $N_2$ | 400 | 2300 |
| $N_2+H_2O$ | 400 | 2300 |

Since the emissions from $N_2$ I molecules in argon plus water and helium plus water plasmas have very weak intensity and the noise to signal ratio is very high, it was not possible to estimate temperatures in those cases.

From the results of temperatures shown in Table 1 and Fig. 3 (a-d), we can see that argon plasma exhibits the higher rotational temperature. On the other hand, the argon plasma has the lowest value of vibrational temperature.

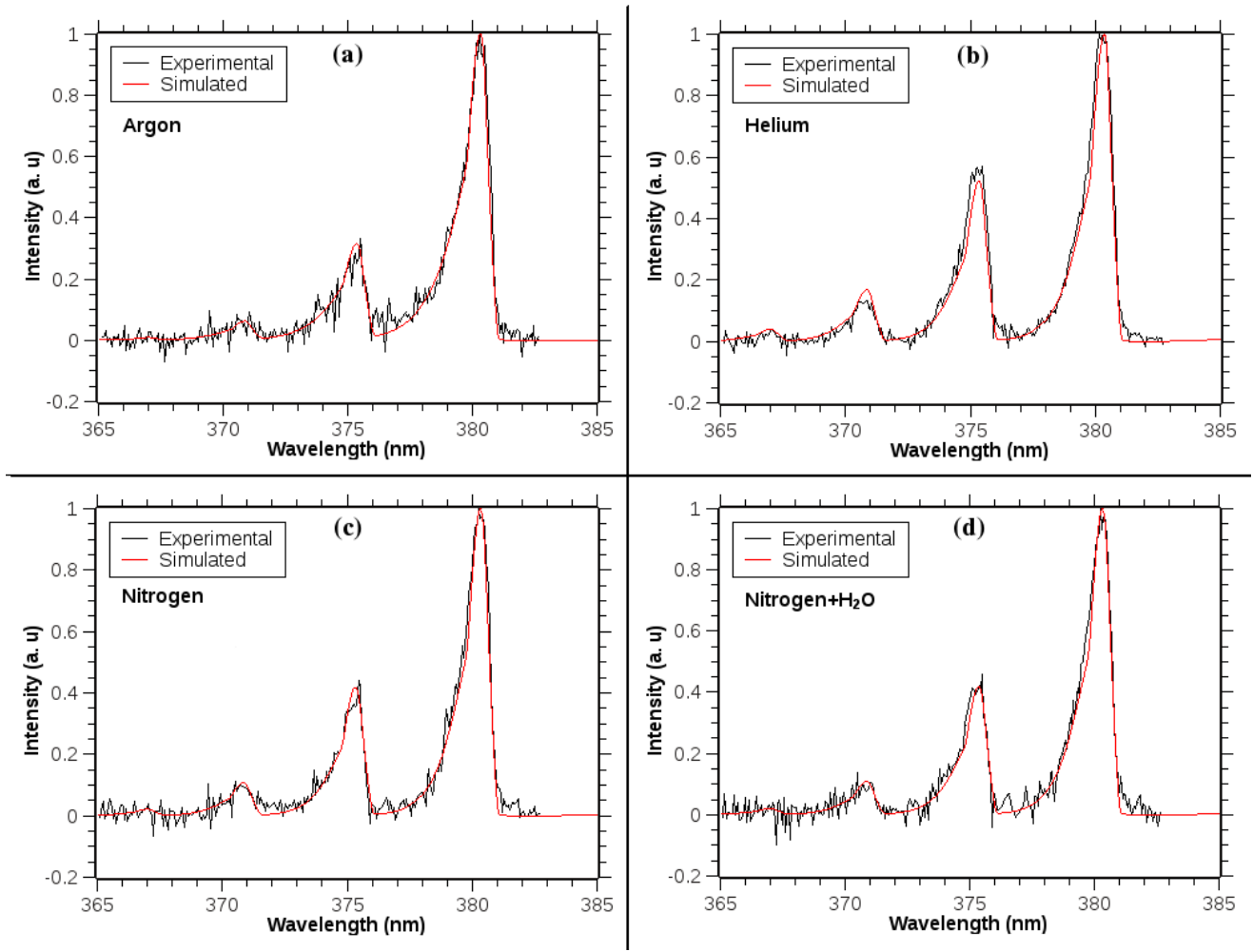

**Figure 3:** Plots of experimental and simulated emission spectrum used to estimate rotational and vibrational temperatures for: (a) Argon, (b) Helium, (c) Nitrogen and (d) Nitrogen plus water vapor. The curves are normalized to 1.

**– Plasma power**

The plasma power estimates were made using a well known method [Ashpis2012] measuring simultaneously, in one plasma pulse, the voltage applied on the electrode and the current across a shunt resistor. Table 2 summarizes the results obtained for plasma power measurements when operating at 6 plasma pulses per second.

**Table 2:** Peak voltage applied to the electrode and power delivered to each kind of plasma

| Gas | Voltage (kV) | Power (mW) |
|---|---|---|
| Ar | 14.0 | 12.8 ± 1.0 |
| Ar+$H_2O$ | 14.5 | 4.1 ± 0.3 |
| He | 15.0 | 7.8 ± 0.6 |
| He+$H_2O$ | 14.5 | 4.2 ± 0.3 |
| $N_2$ | 30.0 | 10.6 ± 1.1 |
| $N_2$+$H_2O$ | 31.5 | 10.6 ± 1.0 |

Looking at Table 2 we can see that the use of argon gas provides the best relationship between applied voltage and power delivered to the plasma. It is followed by helium, nitrogen and nitrogen plus water. And the worst result is for helium plus water and argon plus water. In Table 2 we can also see that the use of water vapor with a gas causes reduction in the power delivered to the plasma, but it is not significative when using nitrogen.

*Treatment of PDMS surfaces*

– **Measurements of water contact angle**

In order to know the effect of plasma treatments on modification of water contact angle of PDMS surfaces we firstly measured the WCA of PDMS samples without plasma treatment. The other measurements of WCA were performed immediately after the surface treatment. In order to estimate WCA we used a commercial camera and the ImageJ2 software [refImageJ]. Figure 4-a shows the variation of WCA in function of plasma treatment time for all kind of plasma used in this work. The frequency of plasma pulses used to treat the samples was 60 Hz. Figure 4-b shows the recovery of WCA in function of aging time for a PDMS sample treated with helium plasma for 90 seconds with 60 plasma pulses per second.

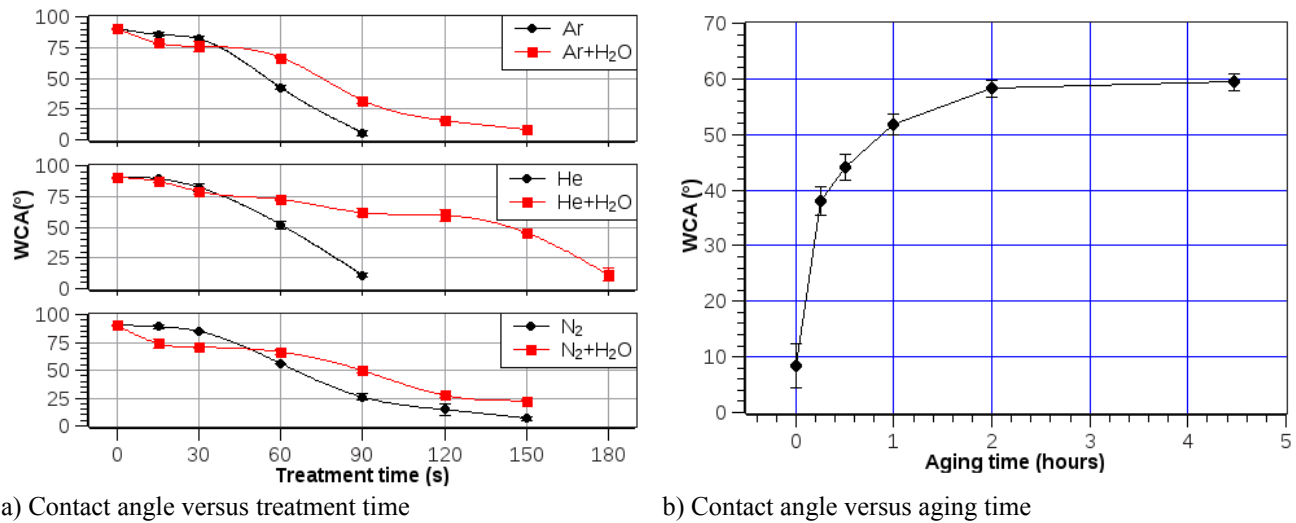

a) Contact angle versus treatment time      b) Contact angle versus aging time

**Figure 4:** (a) Variation of water contact angle in function of the time of surface treatment with different plasmas. Surfaces were treated applying 60 plasma pulses per second. (b) Variation of water contact angle in function of aging time for a PDMS sample treated with helium plasma for 90 seconds. Some error bars for the measurements are too small to appear in the graphs.

As we can see in Fig. 4-a, for short treatment time intervals the use of water slightly improves the WCA; this is more evident when comparing $N_2$ and $N_2+H_2O$ curves. This occurs due to the formation of a thin film of water on the PDMS surface. This thin film of water is visible to naked eye depending on de operation conditions of the DBD device. But the WCA improvement due to this thin film disappears for longer treatment times and the result is worse when using water.

Fig. 4-a shows also the differences in the variation of WCA in function of the time of surface treatment for the gases used. The use of helium and argon gives the fastest reductions of WCA to a very small value (~5-10°), requiring only 90 seconds of treatment at 60 plasma pulses per second (a total of 5400 plasma pulses, approximately) to do the work. For longer treatment times, or a large number of plasma pulses, plasmas of all gases can be used to reduce the WCA to a small value below 10°. However, the use of plasma of $N_2+H_2O$ does not seem to reduce the WCA to the lowest possible value such as the others.

In figure 4-b we can see that the process of recovery of WCA begins in a few minutes.

– **Adhesion tests between PDMS samples**

In order to test the pressure supported in adhesion between two PDMS samples, both samples were exposed to plasma treatment for the same time interval. The two parts are joined together immediately after treatment and were allowed to cure for two days at room temperature. After this period, tests were conducted to determine the pressure supported by the adhesion between samples. The scheme adopted for pressure tests is shown in Figure 5-a. The treated samples were pasted with an appropriated glue (that does not affect the adhesion between the PDMS samples) to the heads of screws of 15mm diameter and a force was applied in an attempt to separate the samples.

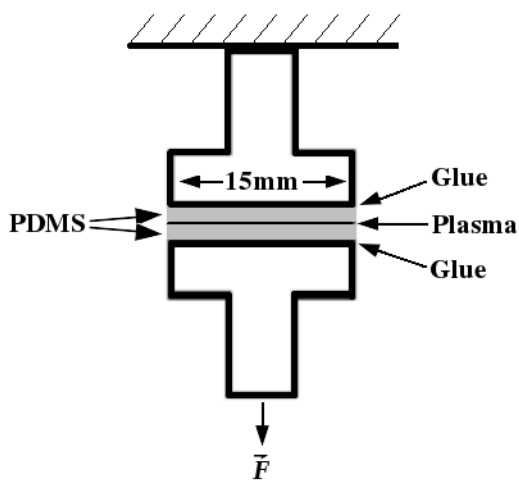
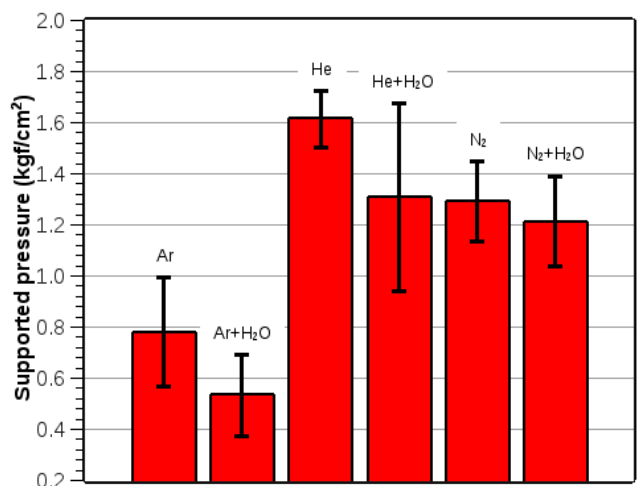

a) Scheme used for pressure tests

b) Pressures supported in adhesion for each kind of treatment

**Figure 5:** a) Transverse section of the scheme used to test the pressure supported between two PDMS samples. b) Supported pressures in adhesion tests between two PDMS samples for each kind of gas used to generate the plasma used in surface treatment. It was applied 20 plasma pulses on each sample of each PDMS pair.

In the cases where the treatment time of the samples were long enough, we verified that the adhesion between PDMS samples are better than adhesion between PDMS and glue, which supported a pressure of ~2.0 kgf/cm² only. This occurred for plasmas of all the gases used for surface treatment. In order to test the differences in the treatment with different gases we performed other tests applying only 20 plasma pulses to each sample of each PDMS pair. The results of these tests are shown in Fig. 5-b.

Figure 5-b shows the averaged (over 5 tests) pressure supported in adhesion between two PDMS samples. Applying only 20 plasma pulses in the PDMS samples we can see that the best result for adhesion was obtained for helium plasma. It is followed by helium plus water, nitrogen and nitrogen plus water plasmas. The argon plus water plasma gives the worst result followed by argon plasma. In general, the treatments with water deteriorated the adhesion results in comparison with the results without water. This fact should be due to the water thin film formation on the PDMS surface. These results can be explained observing the distribution of plasmas over a surface.

In Figure 6 are shown photos of distribution of argon (a) and helium (b) plasmas over surfaces. As can be seen, the distribution of helium plasma over the surface is much more homogeneous than that for argon. The argon plasma forms filaments over the surface and it does not occur with helium plasma, which have a laminar behavior over the surface.

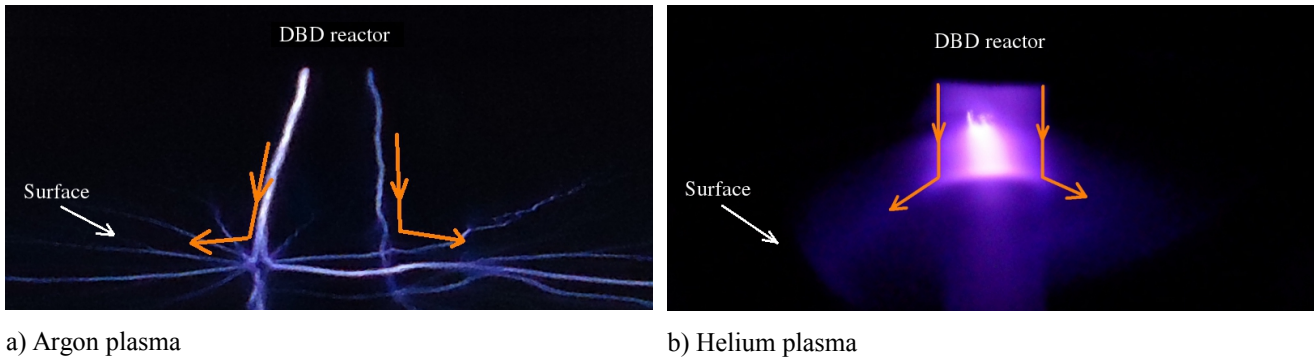

a) Argon plasma  b) Helium plasma

**Figure 6:** Distributions of plasmas over surfaces.

Due to the filamentary behavior of argon plasma over the surface it was not possible to treat the entire surface of a PDMS sample with 15mm diameter applying only 20 plasma pulses. But with helium plasma it looks to be possible. And this is one of the reasons for the result of adhesion using helium to treat the samples to be better than that for argon.

It was not possible to take photos of plasmas formed with the other gases used in this work due to the weak intensity of the light emitted by that plasmas. But in our observations we noted that argon plus water plasma has a surface distribution like the pure argon while all other plasmas have surface distributions like that for helium.

In order to better understand the adhesion improvement after plasma treatments we made AFM analysis of PDMS samples treated with helium plasmas and compared they with a non treated PDMS sample. Figure 7 (a to d) show images of a non-treated PDMS sample (a), a sample treated with 20 plasma pulses (b), a sample treated for 30 seconds operated at 60Hz, ~1800 plasma pulses (c), and a sample treated for 5 minutes at 60Hz, ~18000 plasma puses (d). There were used different PDMS samples to obtain AFM images.

As we can see in Fig. 7, the roughness of PDMS surface changes very much when a few plasma pulses are applied and changes drastically after a long time of plasma treatment. There is a fast reduction in the roughness of the surface. Afret 30 seconds of plasma treatment the roughness is set to its minimum. We also observed that reduction of surface roughness is faster than the reduction of WCA. The calculated root mean square roughness ($R_q$) values are shown in table 3.

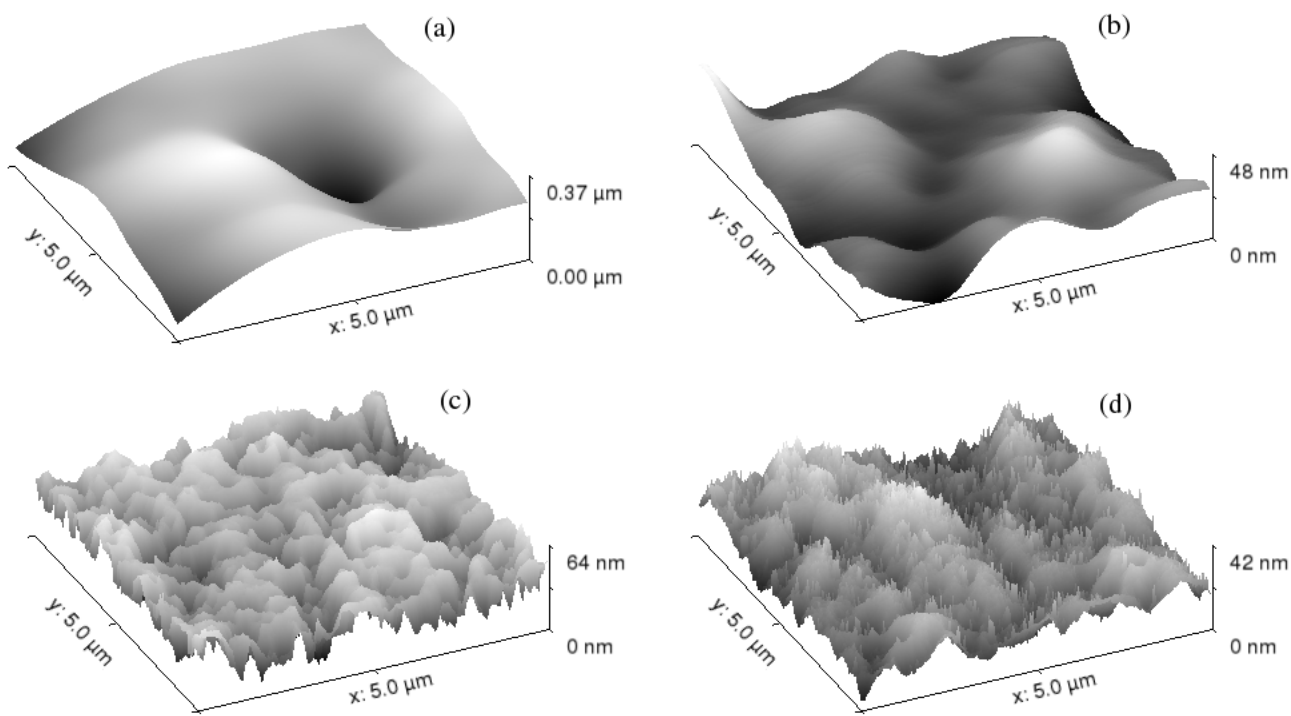

**Figure 7:** AFM images of PDMS samples before and after treatment with helium plasma: (a) sample not treated; (b) sample treated with 20 plasma pulses; (c) sample treated for 30 seconds at 60Hz (~1800 plasma pulses); (d) sample treated for 5 minutes at 60Hz (~18000 plasma puses). Different samples were used to make the AFM images.

**Table 3:** Root mean square roughness values in function of number of plasma pulses applied.

| Number of pulses | $R_q$ (nm) |
|:---:|:---:|
| 0 | 40.95 ± 18.02 |
| 20 | 7.29 ± 2.02 |
| 1800 | 10.03 ± 1.33 |
| 18000 | 5.14 ± 0.81 |

Also in Figure 5, we can see from a macroscopic point of view, the PDMS surface tends to become flat after plasma treatment. From a microscopic point of view, the plasma treatment introduces roughness of small depth and short period. These two last observations are very important in the improvement of adhesion properties because they show an increment in the surface contact area. Both smoothing and introduction of small roughness of the surfaces occurs with other plasma treated polymers [Noeske2004,Nastuta2008,Slepicka2013, Kostov2014]. A reduction in the surface roughness of PDMS after plasma treatment was reported in [Tan2010,Malecha2013].

The surface roughness of the PDMS sample treated for 5 minutes were analyses by AFM in the subsequent days and we did not observed significative modifications with aging time. That is, the change in surface roughness appears to be permanent, in contrast to the reduction of WCA which begins to recover in a few minutes.

## Conclusions

Looking at the emission spectrums of all gases used in this work, the results of pressures supported in adhesion tests and the modification of WCA, we can say that OH molecules are the main species responsible for performing the activation of surfaces in PDMS because they are the reactive species observed in all cases. Excited $N_2$ molecules may also be contributing to the activation of the surface by transferring your vibrational energy.

We observed that there may be a correlation between the vibration temperature and the quality of the adhesion, since the highest pressure supported in the adhesion tests was obtained for the highest vibrational temperature and the lowest pressure supported occurred for the lowest vibrational temperature. We can not observe any influence of the gas temperature in adhesion properties or in the WCA modification. The plasma power also does not seem to cause differences on adhesion properties or WCA modification.

For long treatment times (or a large number of plasma pulses) plasmas of all gases can be used to treat the PDMS surfaces with good results in adhesion and WCA improvements.

For short treatment times (or a small number of plasma pulses) the plasma of helium gas is the best choice for surface treatment due to the homogeneous distribution of the plasma over the surface. And the plasmas of argon and argon plus water are the worst choice due to the filamentary behavior of these plasmas on surfaces. The plasmas of helium plus water, nitrogen and nitrogen plus water, that behave like the helium plasma, can also be used with good results in this case.

Since there is no significative change in the WCA when the PDMS is treated with a few discharges, the modification of WCA after plasma treatment is not the unique mechanism involved in adhesion process. As verified by AFM analysis, the changes in the surface roughness are important for adhesion improvements.

## Acknowledgements

This work was supported by CNPq.